\documentclass[final,authoryear,5p,twocolumn,10pt]{elsarticle}

\usepackage{amssymb}
\usepackage{timet}
\usepackage{graphicx}

\journal{PEPI}

\date{\today}

\newcommand{\vA}{\mbox{\boldmath $A$}}
\newcommand{\vb}{\mbox{\boldmath $b$}}
\newcommand{\vhb}{\hat{\vb}{}}
\newcommand{\vB}{\mbox{\boldmath $B$}}
\newcommand{\vhB}{\,\hat{\!\vB}{}}
\newcommand{\vhA}{\,\hat{\!\vA}{}}

\newcommand{\vJ}{\mbox{\boldmath $J$}}
\newcommand{\vhJ}{\,\hat{\!\vJ}{}}

\newcommand{\vj}{\mbox{\boldmath $j$}}
\newcommand{\vdj}{\mbox{\boldmath $\jmath$}}
\newcommand{\vhj}{\hat{\vdj}{}}

\newcommand{\vr}{\mbox{\boldmath $r$}}
\newcommand{\vu}{\mbox{\boldmath $u$}}
\newcommand{\vv}{\mbox{\boldmath $v$}}

\newcommand{\balf}{\mbox{\boldmath $\alpha$}}
\newcommand{\bbet}{\mbox{\boldmath $\beta$}}

\newcommand{\tauc}{\tau_{\rm c}}

\newcommand{\urms}{u_{\rm r.m.s.}}

\newcommand{\na}{\nabla}

\newcommand{\md}{{\rm d}}
\newcommand{\me}{{\rm e}}
\newcommand{\mf}{{\rm f}}
\newcommand{\mi}{{\rm i}}
\newcommand{\hlf}{{\textstyle\frac{1}{2}}}

\begin{document}

\begin{frontmatter}


\title{Mode analysis of numerical geodynamo models}

\author[1]{M. Schrinner\corref{C}}
\address[1]{MAG (ENS/IPGP), LRA, \'Ecole Normale Sup\'erieure, 
            24 Rue Lhomond, 75252 Paris Cedex 05, France} 
\cortext[C]{corresponding author.} 
\ead{martin@schrinner.eu}    

\author[2]{D. Schmitt}
\address[2]{Max-Planck Institut f\"ur Sonnensystemforschung, 
            Max-Planck-Str. 2, 37191 Katlenburg-Lindau, Germany}     

\author[3]{P. Hoyng}
\address[3]{SRON Netherlands Institute for Space Research, 
            Sorbonnelaan 2, 3584 CA Utrecht, The Netherlands}           


\begin{abstract}
It has been suggested in Hoyng (2009) that dynamo action can be
analysed by expansion of the magnetic field into dynamo modes and
statistical evaluation of the mode coefficients. We here validate
this method by analysing a numerical geodynamo model and comparing
the numerically derived mean mode coefficients with the theoretical
predictions. The model belongs to the class of kinematically stable
dynamos with a dominating axisymmetric, dipolar and non-periodic 
fundamental dynamo mode. The analysis 
requires a number of steps: the computation of the so-called dynamo
coefficients, the derivation of the temporally and azimuthally
averaged dynamo eigenmodes and the decomposition of the magnetic
field of the numerical geodynamo model into the eigenmodes. For
the determination of the theoretical mode excitation levels the
turbulent velocity field needs to be projected on the dynamo
eigenmodes. We compare the theoretically and numerically derived
mean mode coefficients and find reasonably good agreement for
most of the modes. Some deviation might be attributable to the
approximation involved in the theory. Since the dynamo eigenmodes
are not self-adjoint, a spectral interpretation of 
the eigenmodes is not possible.
\end{abstract}


\begin{keyword}
Magnetohydrodynamics \sep Dynamo Theory \sep Geodynamo 
\sep Numerical Simulations \sep Satistical Theory


\end{keyword}

\end{frontmatter}



\section{Introduction}
\label{sec:intro}
The origin of the magnetic field of the Earth is generally understood in 
terms of dynamo action in the liquid outer core. Helical convection and shear 
flows generate and maintain the magnetic field against resistive decay. 
Hydromagnetic simulations confirmed this picture and many groups have 
published numerical geodynamo models in the last 15 years 
\citep[see, e.g.,][]{GR95,KB97,KS97,DO98,RG00,CW07}. 
Even though the computational 
resources do not permit the parameters of the simulations to be Earth-like 
\citep{CA06,CAH10}, the magnetic field of the simulations reproduces many 
features of the observed geomagnetic field. They have also confirmed the idea 
of columnar convection, i.e. that the flow in the core is often organised in 
vortices spiralling parallel to the rotation axis \citep{B75}. 

The simulations opened up the possibility for detailed diagnostics of dynamo 
action. Parameter studies have been undertaken to classify the properties of 
the simulations and to find which parameter combinations produce planetary or 
earth-like dynamos \citep[see, e.g.,][]{KC02,CA06,CAH10}. The mechanisms that 
generate and sustain the magnetic field have also been under scrutiny. 
\citet{KS97} identified elements of the $\alpha\Omega$ mechanism of mean 
field theory, while \citet{OCG99} concluded that in their simulations the 
generation of poloidal and toroidal field resembles the $\alpha^2$ dynamo 
scenario. Another direction of research has been to try and identify the 
mechanisms that induce polarity reversals 
\citep[see, e.g.,][]{SJ99,GC07,DO09,SSCH10a}. Reversals are sometimes 
initiated by a patch of reversed flux appearing deep in the core, which 
subsequently grows and spreads by advection over a larger region and 
eventually dominates over the old polarity field. But no clear picture has 
emerged yet. \citet{AAW08} have developed a new visualisation technique 
(dynamical magnetic field line imaging) and they find that reversals and 
excursions of the dipole axis are associated with what they refer to as 
magnetic upwellings. 

Another benefit of numerical dynamo models is that they permit to measure  
from the simulations the tensors $\alpha_{ij}$ and $\beta_{ijk}$ utilised in 
mean field theory, and \citet{SRSRC07} have computed these mean-field tensors 
with the help of the test field method. Recently, \citet{H09} proposed a 
decomposition of the magnetic field in dynamo eigenmodes and computed the 
statistical properties of the multipole coefficients. The purpose of the 
present paper is to validate this expansion technique and to illustrate some
of the capabilities and problems.
 
\begin{figure*}[t]
\centering\includegraphics[width=15.cm]{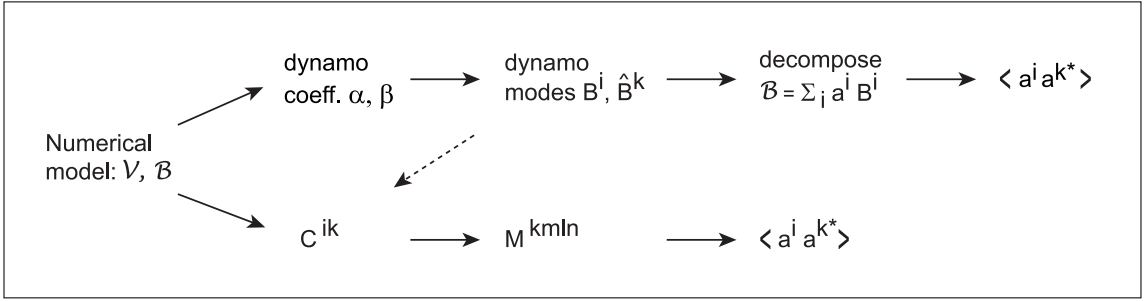}
\caption{Survey of the computations reported in this paper. Starting point is
the numerical dynamo model described in Section~\ref{sec:nummod}. The top line
illustrates how the field ${\cal B}$ is decomposed in dynamo modes. First we
obtain the dynamo coefficients, then the dynamo eigenmodes and their adjoints.
End product are the correlation coefficients $\langle a^ia^{k*}\rangle$. This 
computational train is treated in Section~\ref{sec:modecom}. The correlation 
coefficients may also be computed theoretically, which requires an additional 
computation outlined in the bottom line and described in 
Section~\ref{sec:theormodex}. Measured and theoretical correlation 
coefficients are compared in Section~\ref{sec:modexlev}.
\label{fig:logic}}
\end{figure*}

To this end we decompose the magnetic field of the numerical geodynamo model 
of \citet{OCG99} in dynamo modes and compare the statistical properties of the 
numerically derived mode coefficients with those derived by the theory of 
\citet{H09}. The various steps involved are surveyed in Fig.~\ref{fig:logic}. 
The numerical dynamo model is described in Section~\ref{sec:nummod}. The mode 
decomposition involves various steps presented in Section~\ref{sec:modecom}. 


In Section~\ref{sec:theormodex} we recapitulate the theoretical mode 
excitation levels and in Section~\ref{sec:result} we compare the numerical 
results with the theoretical predictions. Finally, in 
Section~\ref{sec:discsum}, we summarise and discuss our results.

\section{Numerical model}
\label{sec:nummod}
We consider a Boussinesq fluid with electrical conductivity $\sigma$ in a
rotating spherical shell $V$ and solve the momentum equation, the induction
equation and a temperature equation as described in more detail by 
\citet{OCG99}. No-slip mechanical boundary conditions are applied and the 
magnetic field continues outside the fluid shell as a potential field. 
Convection is driven by a temperature difference $\Delta T$ between the inner 
and outer spherical boundaries. The equations are governed by four 
dimensionless parameters, these are the Ekman number $E=\nu/\Omega L^2$, the
(modified) Rayleigh number $Ra=\alpha_T g_0\Delta T L/\nu\Omega$, the Prandtl 
number $Pr=\nu/\kappa$ and the magnetic Prandtl number $Pm=\nu/\eta$. Here,
$\nu$ denotes the kinematic viscosity, $\Omega$ the angular rotation rate, $L$ 
the shell thickness, \(\alpha_T\) the thermal expansion coefficient, $g_0$ the 
gravitational acceleration at the outer boundary, and $\kappa$ is the thermal 
and $\eta=1/\mu\sigma$ the magnetic diffusivity with the magnetic 
permeability $\mu$ and conductivity $\sigma$.

\begin{figure}[h]
\centering{\includegraphics[height=4.cm]{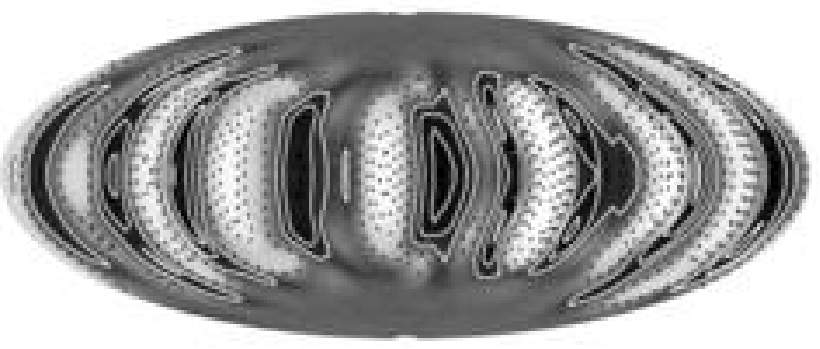}\\[-3.mm]
           \hskip-10.mm
           \includegraphics[width=8.cm]{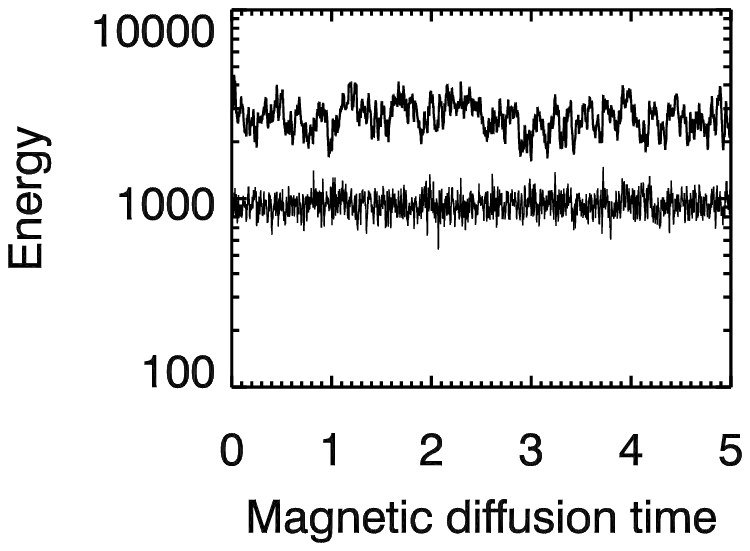}\\[-6.mm]            
          }
\caption{Top: The radial velocity ${\cal V}_{\rm r}$ in the corotating frame 
of the dynamo model under consideration. The snapshot has been taken at 
$r=0.62\,r_0$, where $r_0$ is the outer shell radius. The velocity component 
is normalised to its absolute maximum. Thus, the greyscale coding runs from 
$-1$, white, to $+1$, black, and the contour lines correspond to $\pm 0.1,
\pm 0.3,\pm 0.5, \pm 0.7, \pm 0.9$. Bottom: Magnetic (upper curve) and kinetic 
energy densities of the model versus magnetic diffusion time $L^2/\eta$. 
\label{fig:model}}
\end{figure}

The model under consideration is taken from \citet{OCG99} with $E=10^{-4}$, 
$Ra=334$, $Pm=2$ and $Pr=1$. The resulting convection pattern is strongly 
columnar and symmetric with respect to the equatorial plane, see 
Fig.~\ref{fig:model}. The magnetic field is largely dipolar, strictly 
antisymmetric with respect to the equatorial plane and 
approximately axisymmetric. The energy in the axisymmetric component of 
the field is about 40\% of the total magnetic energy.

The magnetic energy density exceeds the kinetic energy density, as 
illustrated in Fig.~\ref{fig:model}. The magnetic Reynolds number $Rm=\urms 
L/\eta$ is approximately $88$. Here $\urms$ denotes the r.m.s. magnitude of 
the velocity. Although kinetic and magnetic energy densities vary chaotically 
in time, the dipole axis of the magnetic field is stable and reversals do not 
occur. This example belongs to the class of kinematically stable dynamos 
identified by \citet{SSCH10a}, their model 2, which means that there exists 
no magnetic field that will grow exponentially if it is kinematically 
advanced with the time-dependent, saturated, velocity field taken from the 
self-consistent calculation.


\section{Mode decomposition}
\label{sec:modecom}
In principle the magnetic field of a dynamo may be decomposed in any suitable
set of base functions. We could take, for example, the decay modes of a
homogeneous sphere or spherical shell. However, there are reasons to prefer 
another set: the eigenfunctions of the dynamo operator averaged over time and 
azimuth of the numerical model, see \citet{H09} and Section~\ref{sec:result}.
The dynamo operator describes the average dynamo action of a given flow. For 
statistically steady and kinematically stable dynamos, to which we restrict 
ourselves here, the actual magnetic field of the numerical model is expected 
to be well represented by the fundamental eigenmode and a few overtones. Since 
the magnetic field of the numerical dynamo model is largely axisymmetric, we 
restrict the decomposition to the axisymmetric dynamo modes. Likewise we 
ignore symmetric modes since the field of the model is strictly antisymmetric 
with respect to the equatorial plane.

A mode decomposition then requires three steps. The first is that we compute
the dynamo coefficients $\alpha_{ij}$ and $\beta_{ijk}$ of the model. Then we
determine the eigenfunctions and eigenvalues of the dynamo equation, and
finally we decompose the field $\cal{B}$ of the numerical model in these
eigenfunctions.


\subsection{Properties and notations}
\label{sec:sumnot}
Before we engage in details we summarise some of the notations and properties 
that we employ:
\begin{enumerate}[-]
\item
The magnetic field ${\cal B}$, vector potential ${\cal A}$, current 
${\cal J}$ and the velocity ${\cal V}$ in the dynamo region $V$ are written 
in calligraphic font; ${\cal B}=\na\times{\cal A}$, etc. Furthermore, we make 
the usual split in mean flow $\vv$ and turbulent flow $\vu$: ${\cal V}=\vv+
\vu$. 
\item
$\vB$ is the mean field $\langle{\cal B}\rangle$, and $\vB^i$ are the 
eigenfunctions of the dynamo equation of the numerical model, briefly 
referred to as the dynamo modes. We use upper indices to enumerate modes and 
lower indices for vector components.
\item
The mode decomposition makes use of the adjoint eigenfunctions, indicated by a 
hat $\ \hat{}\,$, e.g. $\vhB^k$ is the adjoint of $\vB^k$. 
\item
The adjoint operation and $\na$ commute. The basic entities are the field and 
its adjoint. The adjoint current $\vhJ^k$ is defined as $\na\times\vhB^k$,
likewise if $\vhB=\na\times(\cdots)$, then $(\cdots)=\vhA$.
\item
We absorb the factor $4\pi/c$ in the definition of all currents. So ${\cal J}=
\na\times{\cal B}$, $\,\vhJ^k=\na\times\vhB^k$, etc.
\item
$\vb^i$ are the decay modes of the spherical shell $V$; the decay modes are 
self-adjoint, so we may adopt $\vhb^i=\vb^{i*}$; $\vhj^i=\vj^{i*}=\na\times
\vb^{i*}$, and we shall assume that they are orthonormal:
\begin{equation}
\int_{V+E}\vb^{k*}\cdot\vb^i\,\md^3\vr\,=\,\delta^{ki}\ .
\label{eq:decort}
\end{equation}
Here $E$ is the vacuum exterior to $V$ (not to be confused with the Ekman 
number).
\end{enumerate}


\begin{figure}
\centering\includegraphics[width=6.cm]{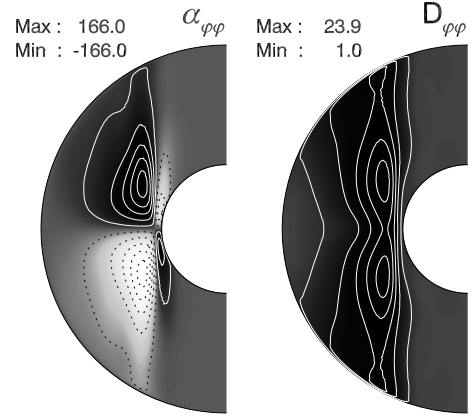}
\caption{Two components of the dynamo coefficients. Left:
$\alpha_{\varphi\varphi}$ in units of $\eta/L$. On the right:
$D_{\varphi\varphi}=\eta+\hlf\beta_{\varphi r\theta}-\hlf\beta_{\varphi \theta
r}$ in units of $\eta$. For each plot the grey scale is separately adjusted to
its maximum modulus with white as negative and black as positive. The contour
lines correspond to $\pm 0.1$, $\pm 0.3$, $\pm 0.5$, $\pm 0.7$, and $\pm 0.9$
of the maximum modulus.
\label{fig:alphabet}}
\end{figure}


\subsection{Dynamo coefficients}
\label{sec:mfcoeff}
The action of the small-scale turbulent flow $\vu$ on the mean magnetic field
is parametrised by the so-called dynamo coefficients, $\balf$ and $\bbet$. 
They are, in general, tensors of second and third rank, respectively. The 
various coefficients describe, e.g., the anisotropic dynamo effects of helical 
flows ($\alpha$-effect), the transport of the mean magnetic field and the 
anisotropic turbulent magnetic diffusivity.

The dynamo coefficients are computed with the help of the test field method
developed by \citet{SRSRC05,SRSRC07}. For a set of suitably chosen test 
fields, the induction equation for the fluctuating field
is solved with the actual flow of the numerical model, the electromotive 
forces are computed, and then averaged in azimuth and time. Finally the 
coefficients are derived from the projection of the mean electromotive forces 
on the corresponding test fields. For details we refer to 
\citet{SRSRC07,S11}. For axisymmetric mean fields the $\balf$-tensor 
has 9 non-zero elements while the $\bbet$-tensor has 18 non-zero entries. As 
an example we show in Fig.~\ref{fig:alphabet} two components for the dynamo 
model under consideration: $\alpha_{\varphi\varphi}$ and $D_{\varphi\varphi}=
\eta+\hlf\beta_{\varphi r\theta}-\hlf\beta_{\varphi\theta r}$, i.e. the  
$\varphi\varphi$-component of the turbulent diffusivity plus the molecular 
diffusivity $\eta$. For more details on the representation of the dynamo 
coefficients we refer to \citet{SRSRC07}, in particular Eqs.~(10)-(13) and 
(15e). The dynamo coefficients reflect the action of the convection columns 
on the mean magnetic field outside the inner core tangent cylinder. They are 
almost zero inside the tangent cylinder where no convection takes place.


\begin{table}
\caption{Eigenvalues of the first 13 antisymmetric and axisymmetric dynamo 
modes in units of $\eta/L^2$, numbered according to decreasing $\Re\lambda$, 
and computed as in \citet{SSJH10b}, with $n_\mathrm{max}=16$, $l_\mathrm{max}
=32.$ For a complex pair of modes, the larger mode number is taken to 
correspond to $\Im\lambda<0$.}
\label{tab:modes}
\vspace{2.mm}
\centering\begin{tabular}{cl}
\hline
mode number $i$ & eigenvalue $\lambda_i$ \\
\hline
$0$             & $(-3.875,\ 0.00)$      \\
$1,2$           & $(-34.83,\ \pm 10.31)$ \\
$3$             & $(-40.63,\ 0.00)$      \\
$4$             & $(-42.88,\ 0.00)$      \\
$5$             & $(-70.13,\ 0.00)$      \\
$6$             & $(-72.25,\ 0.00)$      \\
$7$             & $(-76.44,\ 0.00)$      \\
$8,9$           & $(-98.22,\ \pm 43.87)$ \\
$10$            & $(-106.0,\ 0.00)$      \\
$11$            & $(-135.4,\ 0.00)$      \\
$12,13$         & $(-135.9,\ \pm 9.318)$ \\
\hline
\end{tabular}
\end{table}


\begin{figure}
\centering{\includegraphics[width=7.5cm]{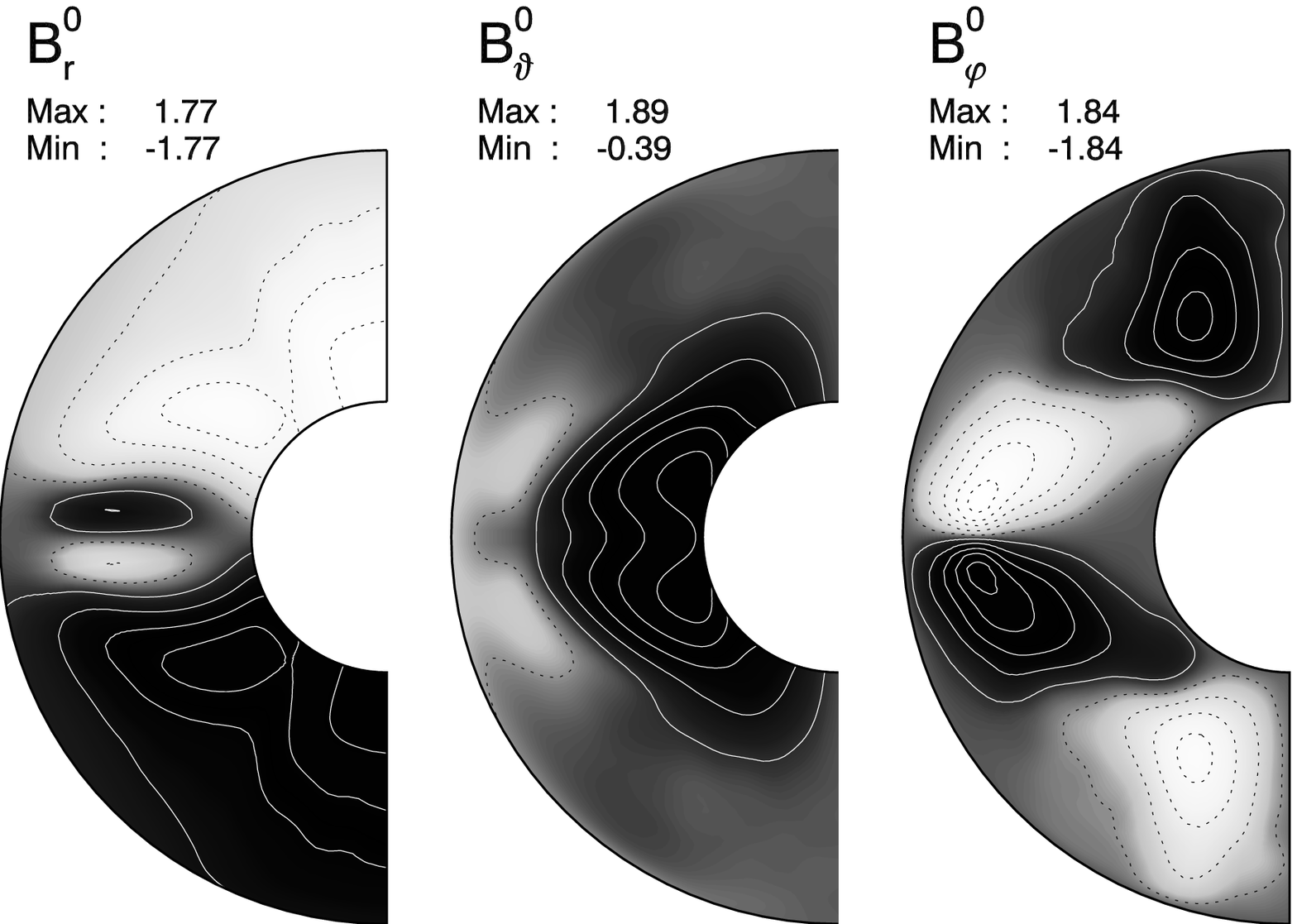}\\[2.mm]
           \includegraphics[width=7.5cm]{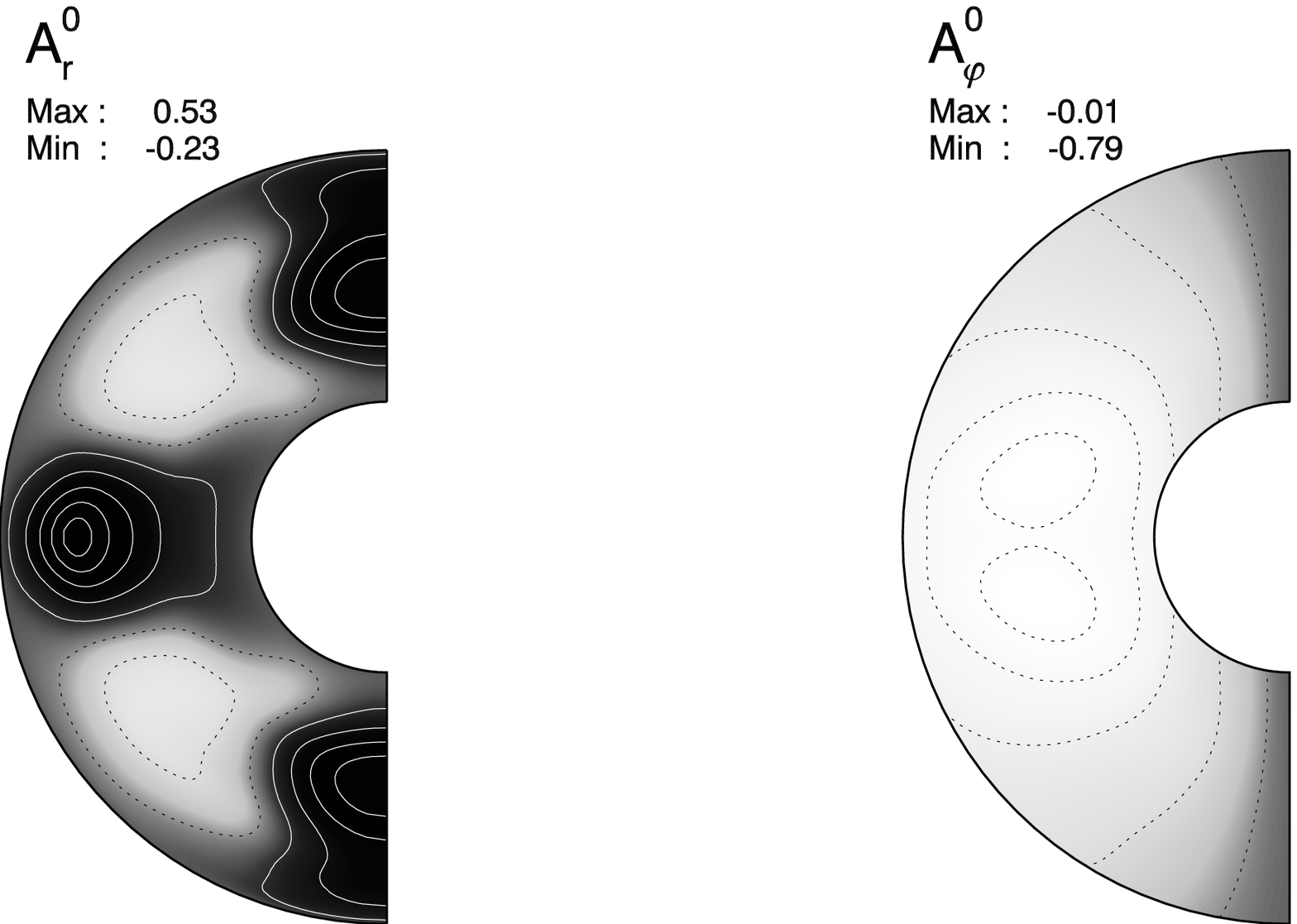}\\[2.mm]
           \includegraphics[width=7.5cm]{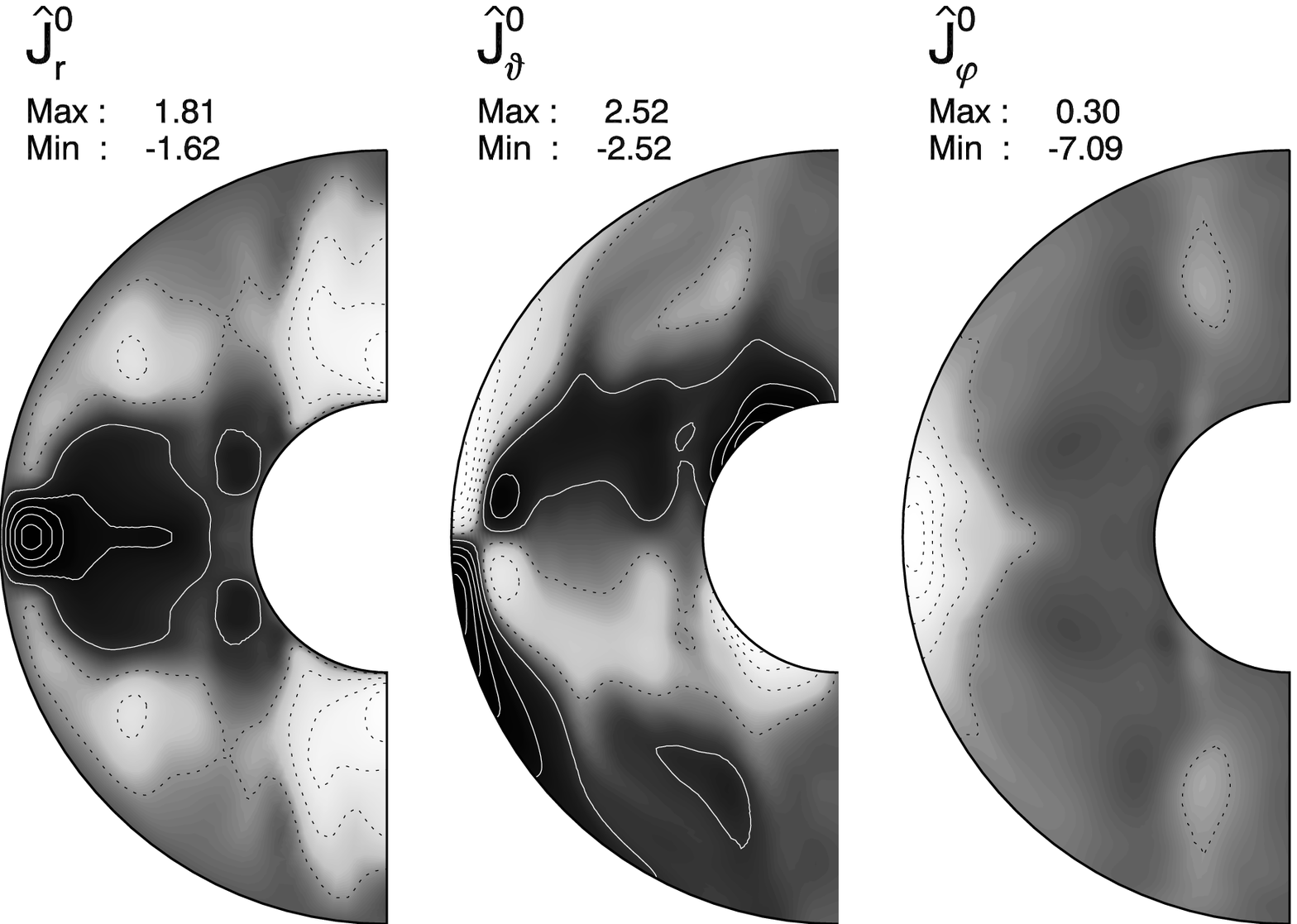}
          }
\caption{The magnetic field $\vB^0$ (top), the vector potential $\vA^0$ 
(middle), and the adjoint current $\vhJ^0$ (bottom) of the fundamental dipole
mode of the numerical model. In the gauge we adopted for the vector potential 
\citep{SSJH10b}, $A^0_\theta$ is zero due to axisymmetry. 
\label{fig:JandA}}
\end{figure}


\subsection{Dynamo modes}
\label{sec:dynmod}
We next solve the eigenvalue problem of the dynamo equation
\begin{equation}
\lambda\vB=\na\times D\vB\ ,
\label{eq:dyn1}
\end{equation}
with
\begin{equation}
D\vB=\vv\times\vB+\balf\cdot\vB-\bbet:(\na\vB)-\eta\na\times\vB\ .
\label{eq:dynop}
\end{equation}
Here $\vv$ is the mean flow and $\vB$ the mean field. The eigenfunctions of 
the dynamo equation are denoted as $\vB^i(\vr)$ and we refer to them as the 
`dynamo modes'. They are constructed by expanding them in the magnetic decay 
modes $\vb^\ell(\vr)$ of the spherical shell $V$:
\begin{equation}
\vB^i=\me^{i\ell}\vb^\ell
\label{eq:expBi}
\end{equation}
Summation convention over double upper indices is adopted. Expansion 
(\ref{eq:expBi}) is now inserted in Eq.~(\ref{eq:dyn1}), after which the 
eigenvalues $\lambda_i$ and the matrix elements $\me^{i\ell}$ may be 
determined as explained in \citet{SSJH10b}. The computation of the matrix 
elements requires that the normalisation of $\vB^i$ is fixed, and we suppose 
that that has been taken care of. Details are given in \ref{sec:appa}. The 
advantage of the new method described by \citet{SSJH10b} is that the velocity 
and the dynamo coefficients do not have to be differentiated, as would be 
necessary in the traditional methods. This is important for the present study, 
as we obtain these coefficients and the velocity from the dynamo code in the 
form of numerical tables.

Since the dynamo modes $\vB^i$ are in general not orthogonal we make use of 
the adjoint modes $\vhB^k$. Together they form a biorthogonal set on $V+E$:
\begin{equation}
\int_{V+E}\vhB^k\cdot\vB^i\,\md^3\vr\,=\,
\int_V\vhJ^k\cdot\vA^i\,\md^3\vr\,=\,\delta^{ki}
\label{eq:bior}
\end{equation}
The second relation follows by integrating by parts, and shows that currents 
and vector potentials also form a biorthogonal set on the volume $V$ of the
dynamo. The adjoint modes may again be found with the help of an expansion in 
decay modes:
\begin{equation}
\vhB^k=\mf^{ks*}\vb^{s*}\ ,
\label{eq:expaBi}
\end{equation}
and \citet{SSJH10b} show that $\mf=(\me^{-1})^\dagger$ where $\dagger$ indicates 
the Hermitean adjoint. For details and the question of gauge invariance we 
refer to this paper. Since the eigenmodes and their adjoints are constructed 
as linear combinations of the decay modes they automatically satisfy the 
boundary conditions, because the decay modes do so. We refer to 
Section~\ref{sec:sumnot} for a summary of the various magnetic field related 
concepts that we have introduced (model field $\cal{B}$, dynamo modes $\vB^i$, 
decay modes $\vb^i$, their adjoints, currents and vector potentials).

As explained above, we restrict the mode decomposition to antisymmetric and 
axisymmetric dynamo modes. The eigenvalues of the first of these modes are 
shown in Table~\ref{tab:modes}. The fundamental mode $\vB^0$ is non-periodic 
and antisymmetric with respect to the equator, and may, on that account, be 
representative for the geodynamo. Non-periodic modes $\vB^i$ are themselves 
real. The structure of the fundamental mode $\vB^0$ together with the 
corresponding vector potential $\vA^0$ and adjoint current $\vhJ^0$ is shown 
in Fig.~\ref{fig:JandA}. 
 
Remarkable is the large drop in growth rate beyond the fundamental mode which
underlines the outstanding role of the fundamental mode in representing the
magnetic field of this dynamo model. For completeness we mention that the
spectrum of the symmetric modes is similar to that of the antisymmetric modes,
but the fundamental symmetric mode decays more strongly, at a rate of
$-6.298\,\eta/L^2$.


\subsection{Field decomposition}
\label{sec:fielddecom}
The magnetic field ${\cal B}$ of the numerical dynamo model
is represented by an expansion in dynamo eigenmodes $\vB^i$:
\begin{equation}
{\cal B}(\vr,t)=a^i(t)\vB^i(\vr)
\label{eq:expB}
\end{equation}
So we first construct the dynamo modes from the decay modes, and then we
represent the field ${\cal B}$ of the dynamo as a superposition of dynamo
modes. Because the dynamo modes (eigenfunctions of the dynamo equation) 
may be supposed to constitute a complete function set they may be used to 
represent {\em any} arbitrary magnetic field. 

\begin{figure}
\centering\includegraphics[width=7.5cm]{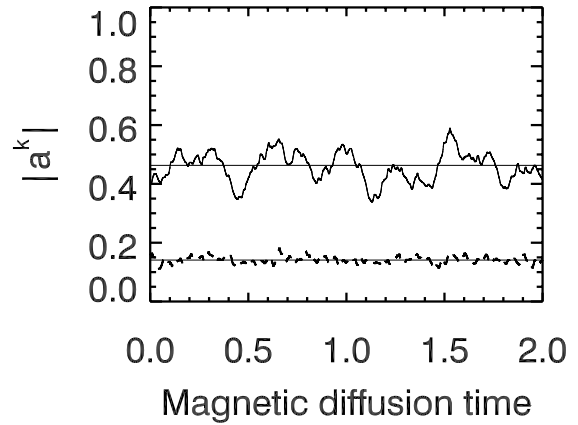}
\caption{Modulus of the amplitude of the fundamental mode ($k=0$, top) and of 
the first overtone ($k=1$, broken line), as a function of time in units of the 
magnetic diffusion time $L^2/\eta$. The horizontal line indicates the mean 
modulus.
\label{fig:msmodc}}
\end{figure}

The expansion coefficients $a^i$ may be computed by taking the inner product
with the adjoint $\vhB^k$ and integrating over all space:
\begin{equation}
a^k\,=\,\int_{V+E}\vhB^k\cdot{\cal B}\;\md^3\vr\,=\,
\int_V\vhJ^k\cdot{\cal A}\;\md^3\vr\ .
\label{eq:expcoef1}
\end{equation}
Since $\vB^0$ is real, the coefficient $a^0$ is real. To evaluate the
integrals in (\ref{eq:expcoef1}) we compute first the adjoint currents
$\vhJ^k=\mf^{k\ell*}\vj^{\ell*}$ on the numerical grid, cf. 
Eq.~(\ref{eq:expaBi}). This needs to be done only once. Next we compute at 
each time step the vector potential ${\cal A}$. The integral $\smallint_V
\vhJ^k\cdot{\cal A}\;\md^3\vr$ in (\ref{eq:expcoef1}) follows by taking the 
inner product and summing over the grid. Because of the infinite volume and 
the power-law scaling of $\vhB^k\cdot{\cal B}$ with radius, it is not a good 
idea to evaluate the other expression $\smallint_{V+E}\vhB^k\cdot{\cal B}\;
\md^3\vr$. Fig.~\ref{fig:msmodc} shows $\vert a^k(t)\vert$ for $k=0$ and 
$k=1$ as a function of time, together with the averages $\langle\vert a^k
\vert\rangle$ (horizontal lines). Once the expansion coefficients have been 
determined the computation of the cross correlation coefficient $\langle 
a^ka^{\ell*}\rangle$ is straightforward. 

Two remarks on the magnitude of the expansion coefficients. The $a^k$ are of 
order unity, and that is a result of scaling: in the dynamo code $\cal{B}$ 
has been scaled to be of order unity. And the normalisation (\ref{eq:normal})
makes that the integral in Eq.~(\ref{eq:expcoef1}) is also of order unity, at 
least for the modes of lower order. The second remark is that by adopting 
the normalisation (\ref{eq:normal}) we ensure that all dynamo modes, if 
excited alone and with the same amplitude (i.e. $a^k=1$ other $a^i=0$), have 
the same total magnetic energy, see \ref{sec:appa}.   


\section{Theoretical mode excitation levels}
\label{sec:theormodex}
The r.m.s. mode coefficients may be found from the evolution equation for the
correlation coefficients derived in \citet{H09}: 
\begin{eqnarray}
&&
\left(\frac{\md}{\md t}-\lambda_k-\lambda_\ell^*\right)
\langle a^ka^{\ell*}\rangle\,=\nonumber \\
&&\qquad\qquad\qquad
\left(M^{km\ell n}+M^{\ell nkm*}\right)\langle a^ma^{n*}\rangle\ .
\qquad
\label{eq:dmakal2}
\end{eqnarray}
(summation over $n,m$, not over $k,\ell$). The matrix elements $M^{km\ell n}$
are defined as integrals over velocity cross correlation functions:
\begin{equation}
M^{km\ell n}\,=\,\int_0^\infty\!\md\tau\,
\langle C^{km}(t)C^{\ell n*}(t-\tau)\rangle\ .
\label{eq:mkmln}
\end{equation}
The functions $C^{km}(t)$ are in turn averages of the turbulent flow $\vu$ 
over the volume of the dynamo weighted by `spatial filters' defined by the 
dynamo modes:
\begin{equation}
C^{k\ell}(t)=-\int_V\;\vu\cdot\vhJ^k\times\vB^\ell\;\md^3\vr\ .
\label{eq:ckl}
\end{equation}
This expression suggests that $C^{k\ell}$ may be interpreted as being 
proportional to the work done per unit time by a part of the Lorentz force 
on the flow $\vu$.  


\begin{figure}
\centering\includegraphics[width=7.cm]{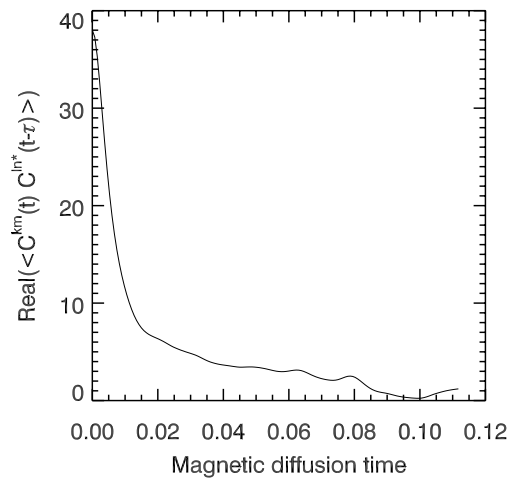}
\caption{The cross correlation function $\langle C^{km}(t)C^{\ell n*}(t-\tau)
\rangle$ as a function of the time shift $\tau$, for $k=\ell=1,\ m=n=0$. The 
vertical axis is in units of $(\nu/L^2)^2=4(\eta/L^2)^2$.  
\label{fig:ckmcln}}
\end{figure}


\subsection{Computation of $M^{km\ell n}$}
\label{sec:compm}
The turbulent flow $\vu$ must be measured and projected on a set of known
vectors $\vhJ^k\times\vB^\ell$, and summed over the spatial grid. This may 
be done on line after each new time step of the dynamo code. The result would 
be a set of time series $C^{k\ell}(t)$. However, the mean flow $\vv$ and 
therefore the turbulent flow $\vu=\cal{V}-\vv$ is not known until the end of 
the simulation, so we must adopt a more elaborate procedure. We project the 
full flow field $\cal{V}$ on $\vhJ^k\times\vB^\ell$, and define 
$\tilde{C}^{km}$ as above but with the full flow field. The required cross 
correlation may now be computed from
\begin{eqnarray}
&&\langle C^{km}(t)C^{\ell n*}(t-\tau)\rangle \nonumber\\[2.mm]
&&\qquad\quad=\langle \tilde{C}^{km}(t)\tilde{C}^{\ell n*}(t-\tau)\rangle
-\langle\tilde{C}^{km}\rangle\langle\tilde{C}^{\ell n*}\rangle\ .
\nonumber \\
\label{eq:CC}
\end{eqnarray}
A result is shown in Fig.~\ref{fig:ckmcln}. The computation of the averages 
$\langle\tilde{C}^{km}\rangle$ in (\ref{eq:CC}) is straightforward, and we 
briefly explain how the cross correlation functions are computed, suppressing 
all details on the implementation in the dynamo code. Writing momentarily 
$x=\tilde{C}^{km}$ and $y=\tilde{C}^{\ell n*}$, we compute, for a given time 
$t$, an array of values $x(t)y(t-i\cdot\Delta\tau)$ for $i=0,..n$. Next we 
select other (later) starting times $t$, $N_{\rm av}$ in total, and average 
$x(t)y(t-i\cdot\Delta\tau)$ over these starting times $t$ for fixed $i$. In 
this way we obtain $\langle x(t)y(t-\tau)\rangle$ for time shifts $\tau$ 
ranging from 0 to $\tau_{\rm cu}=n\Delta\tau$. The upper limit $\tau_{\rm cu}$ 
to the various correlation times $\tauc$ is not known a priori but has to be 
determined empirically from graphs similar to Fig.~\ref{fig:ckmcln}. The 
results presented here have been derived with $\tau_{\rm cu}=0.113\,L^2/\eta$, 
$\Delta\tau=7.5\cdot 10^{-4}\,L^2/\eta$ and $N_{\rm av}=6\cdot 10^4$. Finally 
we determine $M^{km\ell n}$ by an integration over $\tau$, in accordance with 
Eq.~(\ref{eq:mkmln}), carried out in the post-processing.


\section{Results}
\label{sec:result}
We shall now compare the numerical results with the theoretical predictions 
and draw some conclusions.


\subsection{Applicability of the statistical theory}
\label{sec:applic}
The statistical theory that we used here has been formulated by \citet{H09}, 
and the conditions of the applicability of the theory have been worked out in 
Sections V and VI of that paper. The first requirement is that the correlation 
time $\tauc$ is short: 
\begin{equation}
C^{k\ell}\tauc\ll 1\ . 
\label{eq:stat3}
\end{equation}
In \citet{H09} $C^{k\ell}$ could be estimated only approximately, but here we 
can do better with the help of the numerical model. From Fig.~\ref{fig:ckmcln} 
at $\tau=0$ we deduce that $C^{10}\simeq[38\cdot 4\,\allowbreak 
(\eta/L^2)^2]^{1/2}\simeq 12\,\eta/L^2$. The correlation time may be taken as 
the $1/e$ time in Fig.~\ref{fig:ckmcln}: $\tauc\sim 0.01\,L^2/\eta$. In this 
way we obtain:
\begin{equation}
C^{10}\tauc\simeq 0.1\ .
\label{eq:stat4}
\end{equation}
$C^{k\ell}$ can be shown to increase slowly with mode number $k$ as $L/L_k$, 
where $L_k$ is the characteristic length scale in mode $k$ ($L_k<L_0=L$). 

The second requirement is that the unperturbed evolution of the system is 
slow. Since the mean flow $\vv$ is very small, this condition reduces to the 
requirement derived in \citet{H09}, relation (21):
\begin{equation}
\frac{\eta\tauc}{L_kL_\ell}\;\simeq\; 
0.01\;\frac{L^2}{L_kL_\ell}\;\ll\,1\ .
\label{eq:stat5}
\end{equation}
From these results we conclude that the first few dynamo modes of the 
numerical model obey the condition of a short correlation time and of a slow 
unperturbed evolution, though not by a wide margin. We return to this point 
in Section~\ref{sec:modexlev}.


\begin{figure}
\centering\includegraphics[width=9.cm]{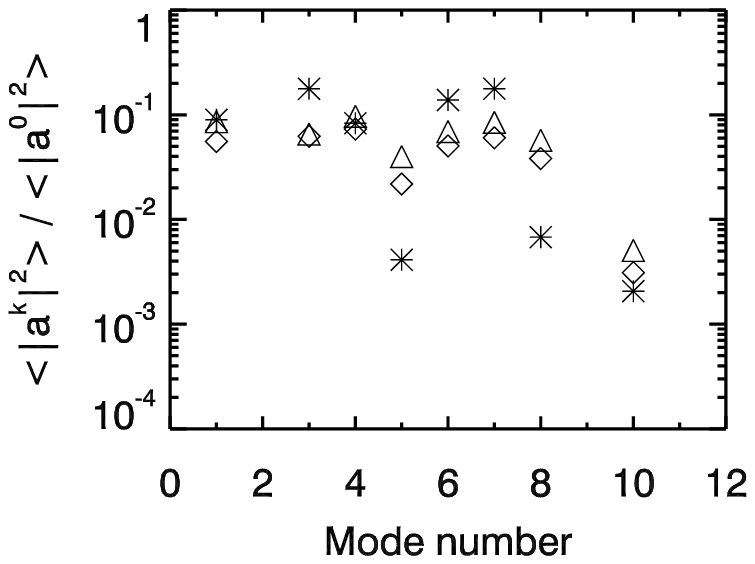}
\caption{Mode excitation levels $\langle\vert a^k\vert^2\rangle$ as a function 
of the mode number defined in Table 1: $\ast=\langle\vert a^k\vert^2\rangle$ 
as measured from the numerical model; $\Diamond\ =\langle\vert a^k\vert^2
\rangle$ computed from (\ref{eq:mak2}); $\triangle\ =\langle\vert a^k\vert^2
\rangle$ computed from (\ref{eq:dmakal2}). The excitation levels have been 
normalised to that of the fundamental mode, whose value is measured to be 
$\langle\vert a^0\vert^2\rangle=0.226$. The levels of mode 1 and 2 are equal 
since they are a complex pair; likewise, the levels of mode 8 and 9 are equal. 
\label{fig:exlev}}
\end{figure}


\begin{figure}
\centering\includegraphics[width=9.cm]{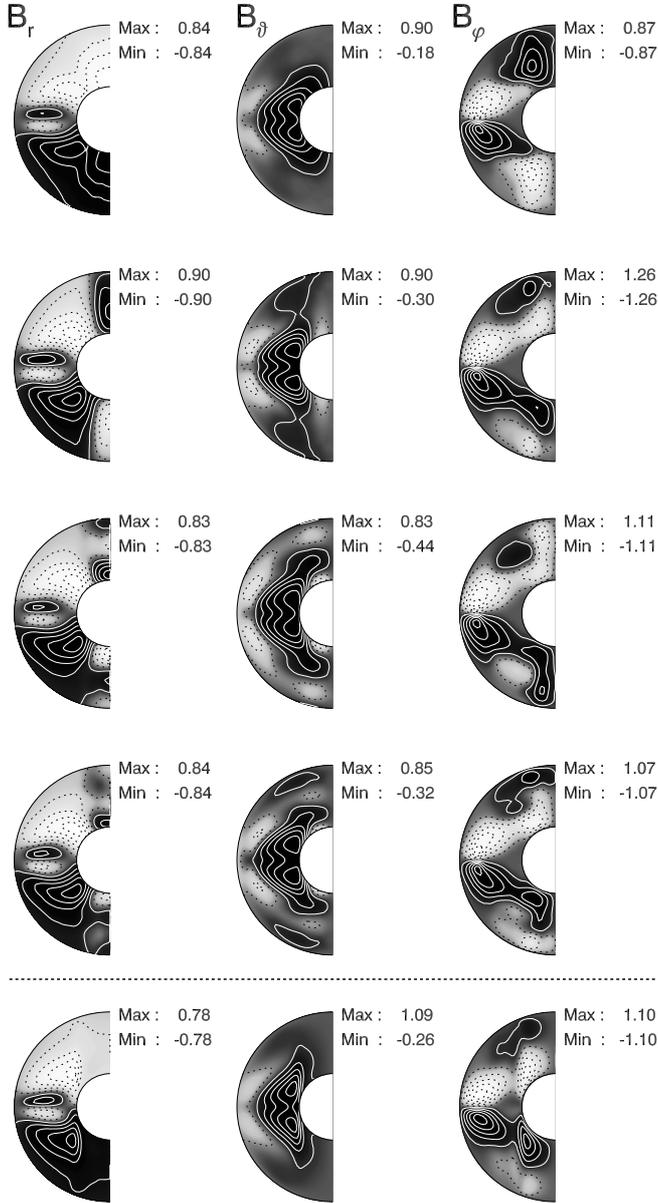}
\caption{The bottom row shows the $r$, $\theta$ and $\varphi$ component of
the dynamo field $\cal{B}$ averaged over azimuth and time. The rows above the horizontal 
line show the field expansion $\Sigma_{k=0}^{k=p}\;a^k\vB^k$ averaged over 
the same period and truncated at various levels. At the top is $p=0$, in the 
downward direction followed by $p=5$, $p=7$ and $p=13$.
\label{fig:eigenmodeapp}}
\end{figure}


\subsection{Mode excitation levels}
\label{sec:modexlev}
We now come to the r.m.s. mode excitation levels. We consider first the 
special case that the fundamental mode is dominant, that is, $\langle\vert 
a^0\vert^2\rangle\ll\langle\vert a^i\vert^2\rangle$ for $i\geq 1$. In that 
case we may ignore all terms in the summation in the right hand side of 
Eq.~(\ref{eq:dmakal2}) except $m=n=0$. We put $\md/\md t=0$ (the dynamo is
supposed to be in a quasi-stationary state) and take $k=\ell$ to
obtain:\footnote{In relation (40) of \citet{H09}, the $\Re$ in front of 
$M^{k0k0}$ has been erroneously omitted.}
\begin{equation}
\langle\vert a^k\vert^2\rangle\ \simeq\ \frac{\Re M^{k0k0}}{-\Re\lambda_k}\;
\langle\vert a^0\vert^2\rangle\ .
\label{eq:mak2}
\end{equation}
Once the $M^{k0k0}$ have been computed as in Section~\ref{sec:compm}, 
relation (\ref{eq:mak2}) allows us to obtain the overtone excitation levels 
relative to that of the fundamental mode. In much the same way we could also 
derive expressions for the cross correlations $\langle a^ma^{n*}\rangle$, 
but we restrict ourselves here to the autocorrelations. The result is plotted 
in Fig.~\ref{fig:exlev} as $\Diamond$, while the measured overtone excitation 
levels are plotted as $\ast$. The agreement between theory and measurement 
seems reasonable to good for all modes except mode $5$, $8$ and $9$. 

As it is not clear to what extent the fundamental mode amplitude is really 
dominant, we have investigated if the solution improves by employing the full 
equation (\ref{eq:dmakal2}). To this end we take again $k=\ell$ and $\md/\md 
t=0$ in Eq.~(\ref{eq:dmakal2}) and evaluate the entire right hand side, the 
$M^{km\ell n}$ as in Section~\ref{sec:compm} and the $\langle a^ma^{n*}
\rangle$ are measured from the numerical data. In the end we divide by 
$-2\Re\lambda_k$. The resulting excitation levels have been plotted as 
${\triangle}$ in Fig.~\ref{fig:exlev}. Since the $\triangle$'s coincide 
closely with the $\Diamond$'s, our tentative conclusion is that the higher 
order terms in the summation in Eq.~(\ref{eq:dmakal2}) may indeed be ignored, 
and that fundamental mode amplitude of the numerical model is in fact 
dominant.

The `flat level' in Fig.~\ref{fig:exlev} may be explained with the help of 
estimate (43) of \citet{H09}: 
\begin{equation}
\frac{\langle\vert a^k\vert^2\rangle}{\langle\vert a^0\vert^2\rangle}\,
\sim\,N^{-1}\sim\,0.1\ ,
\end{equation}
which agrees rather well with Fig.~\ref{fig:exlev}. Here $N$ is the number 
of convective elements, which we estimate to be of order 10 
(Fig.~\ref{fig:model}). So while there is a good agreement at the qualitative 
level, we have no stringent explanation for the small measured excitation 
levels of modes $5$, $8$ and $9$. However, we recall that assumptions
(\ref{eq:stat4}) and (\ref{eq:stat5}) are only marginally fulfilled, 
in particular for higher overtones. Moreover, we may have lost accuracy in
each of the computational steps visualised in Fig.~\ref{fig:logic}.
Notably, the parametrisation of the electromotive force by the dynamo 
coefficients determined and thus the dynamo modes and their eigenvalues will 
be to some extent inaccurate \citep[see also][]{SRSRC07,S11}. 

It is tempting to interprete Fig.~\ref{fig:exlev} as a spectrum, but the 
non-orthogonality of the dynamo modes renders such an interpretation 
impossible, see \ref{sec:appa} for more details.


\subsection{Representation of the field}
\label{sec:goodfit}
The fundamental dynamo mode largely dominates the magnetic field as illustrated 
in Fig.~\ref{fig:eigenmodeapp}. It is obvious that in particular the 
$r$ and $\theta$ components of the time and azimuthally averaged dynamo 
field $\cal{B}$  (bottom row) are fairly well represented by the fundamental 
dynamo mode alone (top row). The $\varphi$ component improves somewhat 
by including a few overtones, but as we do so the $r$ and $\theta$ components 
deteriorate slightly. This problem is fixed again by including a few more 
modes. By the time we have incorporated some $10$ modes the agreement is quite 
good.

The behaviour of the $\varphi$ component shows a peculiarity inherent to an 
expansion in non-orthogonal modes: the approximation is not monotonous. 
Another illustration is the fact that mode 6 and 7 have a large amplitude, see 
Fig.~\ref{fig:exlev}, but they almost cancel each other in the field expansion 
(\ref{eq:expB}), and together they contribute little to the representation of 
the field, as can be seen in the middle row of Fig.~\ref{fig:eigenmodeapp}.


\section{Discussion and summary}
\label{sec:discsum}
Dipolar geodynamo models in the low Rossby number regime are governed by one
fundamental dynamo mode. On average, the fundamental mode has zero growth 
rate, whereas more spatially structured overtones are highly diffusive.  
Small-scale contributions to the magnetic field (i.e. overtones) result
from the deformation of the fundamental mode by the turbulent 
flow; subsequently, they decay due to Ohmic diffusion. The ratio between both 
processes determines their excitation levels. This intuitive picture
proposed and formalised by \cite{H09} is confirmed by the satisfactory
agreement between theoretically predicted and numerically measured
excitation levels in Fig.~\ref{fig:exlev}. Moreover, from a methodological
point of view, the consistency of the whole `train' of computational steps
illustrated in Fig.~\ref{fig:logic} is confirmed.

As noted by \cite{SSCH10a}, the dominance of the fundamental mode in models 
of this dynamo regime is closely related to their time dependence and to
the saturation of the magnetic field. The dipole field in these models is 
stable and polarity reversals are inhibited, because the damping of the 
magnetic field results in the preference of always the same fundamental mode, 
which is continuously quenched and rebuilt. As a by-product, these models may 
be treated kinematically \citep[see][]{SSCH10a} and the mode analysis 
presented in this study is applicable. However, a generalisation of our 
approach to models beyond the kinamatically stable regime would require to 
place velocity modes and magnetic field modes on equal footing 
\citep{CHP10,RB10} and is therefore not easily feasible. 

We expand the magnetic field in dynamo modes which are eigenfunctions of a
non-selfadjoint operator. They form a complete function set but are in general
not orthogonal. This has severe consequences for the interpretation of 
Fig.~\ref{fig:msmodc} or Fig.~\ref{fig:exlev}; they cannot be interpreted as 
spectra. In general, there is no correspondence between the modulus of a mode 
coefficient and the contribution of its associated mode to the dynamo 
field $\cal{B}$. Modes with a comparatively large amplitude may in fact 
contribute only very little. This has been demonstrated in 
Fig.~\ref{fig:eigenmodeapp} for modes 6 and 7 and 
is the  reason for the non-monotonous series approximation (\ref{eq:expB}).  
The ambiguity of mode coefficients is a general problem inherent to most of 
the attempts to model dynamo action by amplitude equations. It could be avoided
by expanding $\cal{B}$ in a orthogonal function set, e.g. the decay modes, 
instead of the dynamo modes. On the other hand, dynamo modes lead to a rather
accurate representation of $\cal{B}$ with a minimum number of modes taken into
account. Furthermore, the compact formulas obtained for 
the excitation levels, (\ref{eq:dmakal2}) and (\ref{eq:mak2}), 
require that $\cal{B}$ is expanded in eigenfunctions of the 
dynamo operator \citep{H09}. 
In short, an expansion in dynamo modes has been preferred because it 
leads to the simplest description and allows for theoretical predictions.  

In this study, we have applied the mode analysis developed 
by \cite{H09} to a geodynamo model in the low Rossby number regime. We find 
satisfactory agreement between theoretical predictions and numerical 
simulations. The expansion of $\cal{B}$ in dynamo modes may serve as a useful 
analytical tool to interpret numerical dynamo models. Besides direct numerical
simulations, low-order models have been successful in illustrating the  
possible interaction of large-scale modes of different symmetries in 
planetary cores \citep[e.g.][]{PFDV09,G10}. The work presented here connects 
both approaches and allows for the possibility to compare predictions 
from low-order models with direct numerical simulations.


\section*{Acknowledgements}
MS is grateful for financial support from the ANR Magnet project. 
The computations have been carried out at the French national computing 
center CINES.

\appendix
\section{Mode normalisation and spectral interpretation}
\label{sec:appa}
First of all, we note that relation (\ref{eq:bior}) does not fix the 
normalisation of the dynamo modes: given a biorthogonal set $\vhB^k, \vB^i$ we 
may construct others, viz.: $\vB^k\rightarrow\mu\vB^k\ ,\quad\vhB^k\rightarrow 
(1/\mu)\vhB^k$, where the factor $\mu$ may depend on the mode number $k$. 
These new sets also obey relation (\ref{eq:bior}). We choose to remove the 
remaining ambiguity, up to an overall phase factor $\exp(\mi\phi)$, by 
imposing that
\begin{equation}
\int_{V+E}\vB^{k*}\cdot\vB^k\,\md^3\vr\,=\,
\int_V\vJ^{k*}\cdot\vA^k\,\md^3\vr\,=\,1\ 
\label{eq:normal}
\end{equation}
(no summation over $k$). The middle relation follows as usual by integrating 
by parts. An immediate consequence of this normalisation is that 
\begin{equation}
(\me\me^\dagger)^{kk}=1\ , 
\label{eq:eedag}
\end{equation}
(no summation over $k$). The diagonal elements of $\me\me^\dagger$ are all 
unity. To see this insert relation (\ref{eq:expBi}) in eq.~(\ref{eq:normal}) 
and use the orthogonality of the decay modes.

The normalisation (\ref{eq:normal}) makes sense for two reasons. The first is 
that in the special case of self-adjoint modes we may choose $\vhB^k=
\vB^{k*}$, and then relation (\ref{eq:normal}) follows automatically from 
(\ref{eq:bior}). The second reason is that, loosely speaking, dynamo modes 
with the same amplitude $a^k$ but excited alone, contain the same total 
magnetic energy. We shall now make this statement precise. 

The energy density of the magnetic field may be written as\newline 
$\mathcal{B}\cdot\mathcal{B}=a^ia^{k*}\vB^i\cdot\vB^{k*}$, 
according to Eq.~(\ref{eq:expB}). 
Next, we insert twice relation (\ref{eq:expBi}), integrate over all space, 
and use the orthogonality of the decay modes. Finally we apply relation 
(\ref{eq:eedag}). The result is:
\begin{eqnarray}
\int_{V+E}\mathcal{B}\cdot\mathcal{B}\,\md^3\vr\,=\,\sum_k\,
\vert a^k\vert^2\,+\,\sum_{i\ne k}\;(\me\me^\dagger)^{ik}\,a^ia^{k*}\,.&&
\nonumber \\[-2.mm]
&& 
\label{eq:spectr1}
\end{eqnarray}
This expression shows that if only one mode is excited at unit amplitude, 
$a^j=1$ and all other $a^k$ zero, then the total magnetic energy 
$\smallint_{V+E}\,\mathcal{B}\cdot\mathcal{B}\,\md^3\vr$ is also unity, 
independent of the mode number $j$.  

Now if the dynamo modes were self-adjoint, then Eq.~(\ref{eq:bior}) becomes $\smallint_{V+E}\vB^{k*}\cdot\vB^i\,\md^3\vr\,=\,\delta^{ki}$. Upon 
inserting again relation (\ref{eq:expBi}) we get $\me\me^\dagger={\rm I}$, 
i.e. the matrix $\me$ is now unitary. In that case the second term in 
(\ref{eq:spectr1}) drops out, and we are left with $\smallint_{V+E}\,
\mathcal{B}\cdot\mathcal{B}\,\md^3\vr=\sum_k\,\vert a^k\vert^2$. And that 
leads straight away to a spectral interpretation of Fig.~\ref{fig:exlev}, 
because the mode number scale on the horizontal axis can be converted into 
a wave number scale. But in general we are stuck with Eq.~(\ref{eq:spectr1}) 
as a whole, and contributions to the magnetic energy remain stored in the 
nonzero mode overlaps $\smallint_{V+E}\vB^{k*}\cdot\vB^i\,\md^3\vr$.


\end{document}